\documentclass[prx,twocolumn]{revtex4-2}
\usepackage{graphicx}
\usepackage{amsmath}
\usepackage{natbib}
\usepackage{epstopdf}
\usepackage{xcolor}

\newcommand{\n}{\nonumber}
\newcommand{\bn}{\begin{eqnarray}}
\newcommand{\en}{\end{eqnarray}}
\newcommand{\eml}{\end{multline}}
\newcommand{\bml}{\begin{multline}}
\newcommand{\h}{\hspace}

\newcommand{\op}[1]{\hat{#1}}

\begin{document}

\title{Entangled Collective Spin States of Two Species Ultracold atoms in a Ring}

 \author{Tom\'a\v{s} Opatrn\'y $^1$ and Kunal K. Das$^{2,3}$}
 \affiliation{$^1$Department of Optics, Palack\'{y} University, 771 46 Olomouc, Czech Republic}
 \affiliation{$^2$Department of Physical Sciences, Kutztown University of Pennsylvania, Kutztown, Pennsylvania 19530, USA}
  \affiliation{$^3$Department of Physics and Astronomy, Stony Brook University, New York 11794-3800, USA}
\begin{abstract}
We study the general quantum Hamiltonian that can be realized with two species of mutually interacting degenerate ultracold atoms in a ring-shaped trap, with the options of rotation and an azimuthal lattice. We examine the spectrum and the states with a collective spin picture in a Dicke state basis. The system can generate states with a high degree of entanglement gauged by the von Neumann entropy. The Hamiltonian has two components, a linear part that can be controlled and switched on via rotation or the azimuthal lattice, and an interaction-dependent quadratic part. Exact solutions are found for the quadratic part for equal strengths of intra-species and the inter-species interactions, but for generally different particle numbers in the two species. The quadratic Hamiltonian has a degenerate  ground state when the two species have unequal number of particles, but non-degenerate when equal.  We determine the impact on the entanglement entropy of deviations from equal particle numbers as well as deviations from the assumption of equal interaction strengths. Limiting cases are shown to display features of a beam-splitter and spin-squeezing that can find utility in interferometry. The density of states for the full Hamiltonian shows features as of phase transition in varying between linear and quadratic limits.
\end{abstract}

\maketitle

\section{Introduction}

Coherent state in a closed loop is a defining paradigm of quantum mechanics, tracing back to De Broglie's explanation of quantization of electronic states in atoms. With the creation of coherence in many body systems, such as with Bose-Einstein condensates (BEC), and progress in trapping them in toroidal configurations, that seminal configuration can be translated to macroscopic scales.  The closed topology and the natural superfluidity associated with degenerate cold gases have focussed most of the interest in this matter on the physics of persistent flows \cite{ramanathan,Phillips_Campbell_superfluid_2013}.  However, the coherent flow in a loop intrinsically comes with angular momentum, and with the circulating modes, parallels can be drawn with states of electrons within atoms, including spin and orbital momenta \cite{Das-Brooks-Brattley}.  The many body nature  \cite{Bloch-RMP-Many-Body} of such macroscopic coherent media and rich nonlinear behavior due to interactions \cite{Das-Huang} means that such ring systems can be a versatile simulator of collective spin states \cite{Kitagawa} and all the rich physics associated with them.  This paper aims to set the basis and framework for such studies.

\begin{figure}[t]
\centering
\includegraphics[width=\columnwidth]{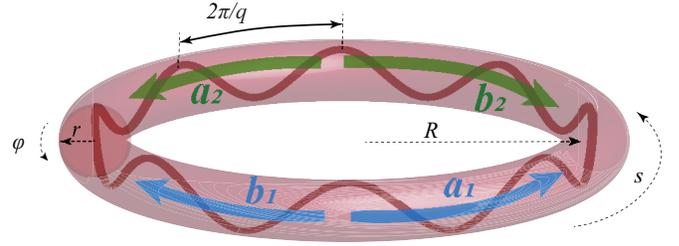}
\caption{(Color online) Two species of atoms labelled $i=1,2$ are trapped in a toroidal trap with the option of an azimuthal lattice potential of period  $2\pi/q$. The two lowest counter-propagating modes for each species are denoted by letters $a,b$. The torus is taken as a wrapped cylinder with our choice of co-ordinates ${\bf r}=(s,r,\varphi)$ shown. }\label{Fig1_schematic}
\end{figure}

Multiple pathways exist for creating ring traps for atoms \cite{ramanathan, Stamper-Kurn-2005,Riis-magnetic,Chapman-magnetic-ring-2001,Boshier-painted-potential,Foot-painted,Lee_Hill_photomask,
Mompart-conical,Dunlop-ring,MEISTER2017375}, some conveniently adaptable to include an azimuthal lattice structures, such as the use of Laguerre-Gaussian (LG) beams \cite{Padgett, Zambrini:07}. While numerous experiments \cite{Campbell_resistive-flow,Phillips_Campbell_superfluid_2013,Phillips_Campbell_hysteresis} have been conducted with cold atoms in ring traps, proportionate effort with the inclusion of lattices are overdue, notwithstanding the rich physics indicated by continuing theoretical works \cite{Aghamalyan-AQUID, Aghamalyan-two-ring-lattice,Piza-ring-lattice,Tiesinga-soliton-lattice,Maik-dipolar,Moreno-Bose-Hubbard,Jezek-Bose-Hubbard-ring-lattice,
Jezek-winding-number,Doron-Cohen-1,Minguzzi-resonant-persistent,Minguzzi-PRA-two-bosons,Penna,Guilleumas-nonlinear_ring, Nigro_2018, Opatrny-Kolar-Das-LMG, Opatrny-Kolar-Das-rotation}.

In previous work, we have shown that a single species in a ring can lead to rich physics: The dynamics can display coherent oscillations between various modes coupled by a lattice \cite{Das-Brooks-Brattley}, nonlinear dynamical behavior like self trapping is evident \cite{Opatrny-Kolar-Das-rotation, Das-Huang},  creation of spin squeezed states and simulation of Lipkin-Meshkov-Glick dynamics are possible \cite{Opatrny-Kolar-Das-LMG}.  However, to examine quantum correlations, associated with multiparticle entanglement that touch on the most intriguing aspects of quantum mechanics, such as EPR and Bell inequalities, that analog in a ring is best implemented with two species of atoms.  Simulation of such intrinsically quantum phenomena with the macroscopic states of a ring motivates this work. Here, we focus on the spectrum and the degree of entanglement of the relevant quantum states in the system, preliminary to examining the dynamics in our continuing work.

In Sec.~II, we present our physical model and derive the two-species Hamiltonian, and transform it to a collective spin description. We set up the states and the measure of entanglement for the system  in Sec.~III. Then in Sec.~IV, we derive analytical expressions for the eigenvalues and for the associated states for the quadratic Hamiltonian that creates entanglement, and we consider various special cases. Section V highlights limiting cases where the system behavior is analogous to a beam-splitter and a spin-squeezer in turn. In Sec.~VI, the density of states for the full Hamiltonian is shown to display features of a phase transition as the Hamiltonian is continuously changed from the linear limit to the quadratic limit. Estimates validating our assumptions along with an outlook of our ongoing work on dynamical applications of these results are presented in our conclusions in Sec.~VII.

\section{Physical Model}
We consider two species of BEC, labelled $j=1,2$ in a toroidal trap as shown in Fig.~\ref{Fig1_schematic}. We take the minor radius to be much smaller than the major radius so that the system can be treated as a cylinder ${\bf r}=(s,r,\phi)$ with periodic boundary condition on $s$.  We assume the confinement along $(r,\phi)$, transverse to the ring circumference to be sufficiently strong to keep the atoms in the ground state $\psi(r,\phi)$ for those degrees of freedom, so that the three-dimensional bosonic field operator can be written in the effective form $ \hat{\Psi}(s)\psi(r,\phi)$.
Integrating out the transverse degrees of freedom, the dynamics can be described by an effective one dimensional Hamiltonian
\begin{eqnarray}
\op{H}&=&\int_0^{2\pi R}\h{-3mm}{\rm d}s\left[{\small\sum_{i=1,2}}\op{\Psi}_i^\dagger
{\left(- \frac{\hbar^2}{2m}\partial^2_s+U_i+\frac{g_i}{4\pi l^2} \op{\Psi}_i^\dagger\op{\Psi}_i\right)\op{\Psi}_i}\right.\n\\
&&\left.\h{3cm}{+\frac{g_{12}}{2\pi l^2} \op{\Psi}_1^\dagger\op{\Psi}_2^\dagger\op{\Psi}_1\op{\Psi}_2}\right].
\label{QF-Hamiltonian}
\end{eqnarray}
where  $g_\alpha=4\pi\hbar^2a_\alpha/m$ is the interaction strength defined by the $s$-wave scattering length $a_\alpha$, with $\alpha \in\{1,2,12\}$  and $l$ is the average harmonic oscillator length for the transverse confinement, assumed to be same for both species. The potential along the ring is taken to be a periodic lattice, rotating with frequency $\omega$, with same period for both.  We take the strength of the potential to be species-selective, indexed by $j=1,2$
\begin{eqnarray}
U_i(s,t)&=&\hbar u_{xi} \cos\left[2q ({\textstyle\frac{s}{R}} - \omega_i t) \right]\n\\&&+ \hbar u_{yi} \sin\left[2q ({\textstyle\frac{s}{R}} - \omega_i t) \right],
\end{eqnarray}
where we allow for potential components symmetric (x) and antisymmetric (y) relative to the co-ordinate origin.

We assume two circulating modes for each species, clockwise and counterclockwise, with field amplitudes $\hat{a}_j,
\hat{b}_j$ which satisfy the bosonic commutator rules.  field operator in the eigenstates of the ring (they would be same for both species since the period is the same)
\bn \op{\Psi}_i(s)=\op{a}_{i}\psi(s)+\op{b}_{i}\psi^*(s) ;\h{3mm} \psi_{n}(s)=\frac{1}{\sqrt{2\pi R}}e^{in(s/R)},\en
We the redefine the operators by replacing $\op{a}_{n}(t)\rightarrow \op{a}_{n}(t)e^{-in\omega t}$ and simplify the notation by defining the effective 1D interaction strengths $\chi_\alpha=\frac{g_\alpha}{4\hbar\pi^2 l^2R} $, unperturbed eigenenergies $\hbar\omega_n=\frac{\hbar^2 n^2}{2mR^2}$ and potential amplitudes $u_{i\pm}=\frac{1}{2}(u_{xi}\pm iu_{yi})$.
\begin{widetext}
\begin{eqnarray}
\op{H}&=&\sum_{i=1,2} \left[ -\hbar q\omega_i(\op{a}_{i}^{\dag}\op{a}_{i} - \op{b}_{i}^{\dag}\op{b}_{i} )
+\hbar\left(u_{i-}\op{a}_{i}^{\dag}\op{b}_{i} +u_{i+}\op{b}_{i}^{\dag}\op{a}_{i} \right)
+{\textstyle\frac{1}{2}}\hbar\chi_i \left( \op{a}_{i}^{\dag}\op{a}_{i}^{\dag} \op{a}_{i} \op{a}_{i}
+4\op{a}_{i}^{\dag}\op{b}_{i}^{\dag} \op{a}_{i} \op{b}_{i}
+\op{b}_{i}^{\dag}\op{b}_{i}^{\dag} \op{b}_{i} \op{b}_{i}\right)\right]
 \nonumber \\
& &  \hbar \chi_{12}\left[ \op{a}_{1}^{\dag}\op{a}_{2}^{\dag} \op{a}_{1} \op{a}_{2}
+\op{a}_{1}^{\dag}\op{b}_{2}^{\dag} \op{a}_{1} \op{b}_{2}
+\op{a}_{1}^{\dag}\op{b}_{2}^{\dag} \op{b}_{1} \op{a}_{2}
+\op{b}_{1}^{\dag}\op{a}_{2}^{\dag} \op{a}_{1} \op{b}_{2}
+\op{b}_{1}^{\dag}\op{a}_{2}^{\dag} \op{b}_{1} \op{a}_{2}
+\op{b}_{1}^{\dag}\op{b}_{2}^{\dag} \op{b}_{1} \op{b}_{2}\right]
\label{Hama}
\end{eqnarray}
\end{widetext}

In order to continue the analysis, we recast the Hamiltonian in terms of the collective spin operators
\begin{eqnarray}
\op{J}_{xi} &\equiv& \frac{1}{2}\left(\op{a}_{i}^{\dag}\op{b}_{i} + \op{a}_{i} \op{b}_{i}^{\dag}\right) , \n\\
\op{J}_{yi} &\equiv& \frac{1}{2i}\left(\op{a}_{i}^{\dag}\op{b}_{i} - \op{a}_{i} \op{b}_{i}^{\dag}\right) , \n\\
\op{J}_{zi} &\equiv& \frac{1}{2}\left(\op{a}_{i}^{\dag}\op{a}_{i} - \op{b}_{i}^{\dag} \op{b}_{i} \right) ,
\end{eqnarray}
so that the Hamiltonian takes the form
\begin{eqnarray}
&&\op{H}=\sum_{i=1,2} \left[-2\hbar q\omega_i  \op{J}_{zi}
+ \hbar u_{xi} \op{J}_{xi} +  \hbar u_{yi} \op{J}_{yi} \right]
\\
\h{-5mm}& &\h{-5mm}
+\hbar \sum_{i=1,2} \chi_i\left[\op{J}^{2}_{xi}+\op{J}^{2}_{yi}\right] + 2 \hbar \chi_{12}\left[ \op{J}_{x1}  \op{J}_{x2} + \op{J}_{y1} \op{J}_{y2} \right]\n
\end{eqnarray}
This left out $2(\hbar\chi_{12}N_1N_2)+2\hbar\chi_i[\op{J}_i^2-N_i]$ where $\op{J}_i^2=\op{J}_{xi}^2+\op{J}_{yi}^2+\op{J}_{zi}^2$ is the total spin operator for each species associated with the eigenvalues $j_i(j_i+1)$.  All of these terms commute with the Hamiltonian and would be conserved in an evolution. \emph{In all our numerical simulations, we will assume natural units, setting $\hbar=m=1$}.

If the intra and inter species couplings are identical, $\chi_1=\chi_2=\chi_{12}=\chi$, which can true to a good approximation for example for Rubidium-87 atoms \cite{Egorov_Rb}, we can express the Hamiltonian as the sum of linear and quadratic parts $\op{H}=\op{H}_L+\chi \op{H}_Q$
\bn \op{H}_L&=&\sum_{i=1,2} \left[-2 q\omega_i  \op{J}_{zi}
+ u_{xi} \op{J}_{xi} +  u_{yi} \op{J}_{yi}\right]\n\\
\op{H}_Q &=&  \left(\op{J}_{x1}+\op{J}_{x2}\right)^2 + \left(\op{J}_{y1}+\op{J}_{y2}\right)^2\label{special_Hamiltonian}\en
neglecting some constants that do not affect the dynamics.  The linear part simply rotates states on the Bloch sphere.  We define the collective operators $ \op{J}_{p\pm}=\op{J}_{p1}\pm\op{J}_{p2}$, with $p\in\{x,y,z\}$ so the quadratic part simply becomes $\op{H}_Q = \op{J}_{x+}^2+\op{J}_{y+}^2$.  The quadratic part is of more significance because it changes the shape of the states, and we will focus on that. In addition to $N_1$, $N_2$, the quadratic part also clearly commutes with  $ \op{J}_{z+}\equiv \op{J}_{z1}+\op{J}_{z2}$.

\begin{figure}[t]
\centering
\includegraphics[width=\columnwidth]{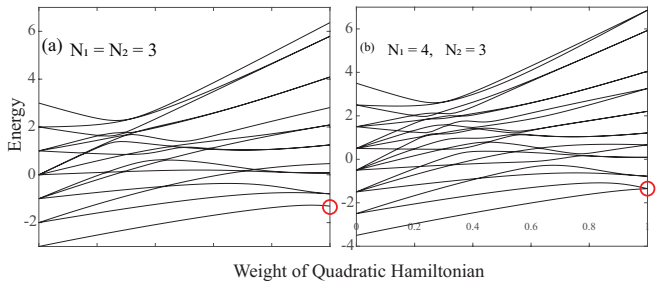}
\caption{(Color online) Plots of the energy as the weight of quadratic Hamiltonian is varied form the linear limit to the quadratic limit.   In the quadratic limit  the general pattern holds as shown, the ground state is (a) non-degenerate when $N_1=N_2$ and (b) degenerate when $N_1\neq N_2$. In the linear limit, we use $\op{H}_L= (\op{J}_{z1}-\op{J}_{z2})$ in Eq.~(\ref{special_Hamiltonian}), corresponding to a rotating ring with no azimuthal lattice. }\label{Fig2_linear_to_quad}
\end{figure}

The linear part of the Hamiltonian can be controlled and even completely turned off with the lattice strength and the rotation, whereas the quadratic part can be likewise controlled or made to vanish with the interaction induced nonlinearity. Thus, in an experiment, it would be convenient to initialize the system in an eigenstate of the linear Hamiltonian.  Thereafter, the components of the linear part of the Hamiltonian can be ramped down and the quadratic Hamiltonian ramped up. In an adiabatic process, the variation of the spectrum would govern the dynamics.  We plot that variation with $(1-w)\op{H}_L+wH_Q$ in Fig.~\ref{Fig2_linear_to_quad}.  Referring to Eq.~(\ref{special_Hamiltonian}), in the plot we absorb the coefficients of the operators $\op{J}_{x/y/z}$ as part of the weight $(1-w)$ of $\op{H}_L$ and the quadratic part is scaled $\op{H}_Q\rightarrow 2\op{H}_Q/(N_1+N_2)$. We choose the linear part of the Hamiltonian with $u_{xi}, u_{yi}\rightarrow 0$, varying $\omega$. The ground state is found to have two distinct behavior.  For $N_1=N_2$, the ground state remains non-degenerate from purely linear to purely quadratic, where as for $N_1\neq N_2$ at the quadratic limit, the ground state is always double degenerate. However, when linear limit has co-propagating modes in the two species, gap may close before reaching the quadratic limit. Still, the state can be initially prepared to sustain the gap so that almost total adiabatic transfer can be achieved from the ground state of the linear Hamiltonian to that of the quadratic Hamiltonian for systems with equal number of particles of both species.

\begin{figure}[t]
\centering
\includegraphics[width=\columnwidth]{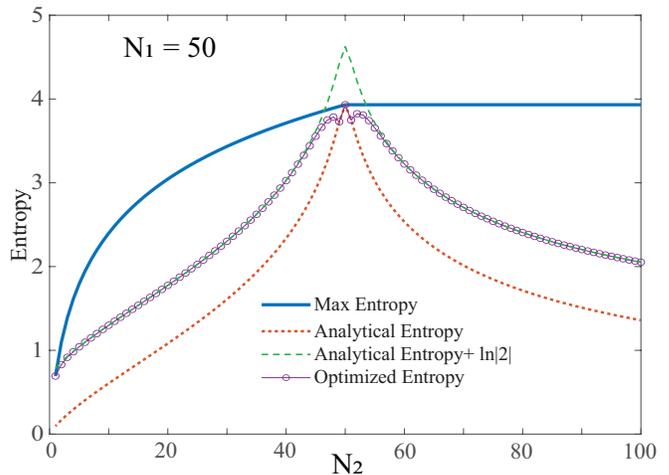}
\caption{(Color online) The effect of imbalance in particle number of the two species is illustrated for the ground state of the quadratic Hamiltonian $\op{H}_Q$, by plotting the associated entropy of entanglement as a function of particle number $N_2$ of the second species, with that of he first one fixed at $N_1=50$. The maximum entropy is set by that of the lower particle count. The dotted red line is computed analytically from the exact ground state in Eq.~(\ref{exact_gs}); the dashed green line has ${\rm ln}|2|$ added to account for the two fold degeneracy, which however is an overestimate close $N_1=N_2$. The circle markers are numerical calculation for an optimal superposition of the degenerate states in Eq.~(\ref{exact_gs}).}
\label{Fig3_imbalance}
\end{figure}

\section{States and Entanglement Entropy}

The system can be described in Fock basis, that specifies the occupation of each of the four modes.  $|n_{a1},n_{b1}\rangle \otimes |n_{a2},n_{b2}\rangle$.  More specifically, we can write the basis as a direct product of Dicke states, the collective spin analog of Fock states, of the two species $|j_1,m_1\rangle\otimes|j_2,m_2\rangle$.  For fixed particle number, we have $j_i=N_i/2$.  The second quantum number specifies eigenstates of \bn \op{J}_{zi}|j_i,m_i\rangle=m_i|j_i,m_i\rangle,\h{3mm} m_i={\textstyle-\frac{N_i}{2},-\frac{N_i}{2}+1,\cdots \frac{N_i}{2}}.\en
We can further simplify to a basis of eigenstates of $ \op{J}_{z\pm}$ that we denote by
\bn \op{J}_{z\pm}|z_+, z_-\rangle =z_\pm |z_+, z_-\rangle.\en
Since $z+$ is a conserved quantum number for our Hamiltonian, we can consider subspaces of fixed $z_+$ independently within which the states are uniquely labelled by a single quantum number  $z_-$:
\begin{eqnarray}\label{numbers}
n_{a1} &=&  {\textstyle \frac{1}{2}}(N_1+z_++z_-), \n\\
n_{b1} &=& {\textstyle \frac{1}{2}}(N_1-z_+-z_-), \n\\
n_{a2} &=&  {\textstyle \frac{1}{2}}(N_2+z_+-z_-),  \n\\
n_{b2} &=&  {\textstyle \frac{1}{2}}(N_2-z_++z_-).
\end{eqnarray}

\begin{figure*}[t]
\centering
\includegraphics[width=\textwidth]{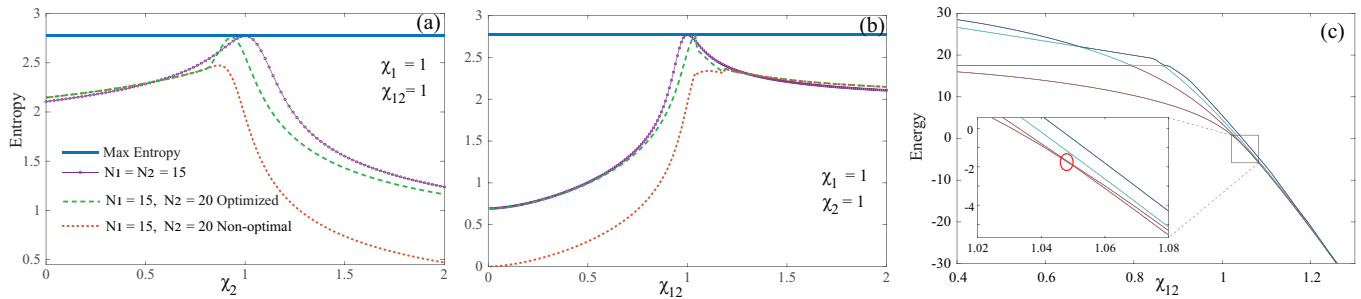}
\caption{(Color online) The entanglement entropy of the ground state of the quadratic Hamiltonian is seen to be maximized when all the interaction strengths are the same $\chi_1=\chi_2=\chi_{12}$ as assumed in Eq.~(\ref{special_Hamiltonian}).  (a) The rate of decline with deviation from that is faster at (a) larger values of an intra-species  $\chi_2$  and (b) smaller values of inter-species interactions $\chi_{12}$.  For $N_1\neq N_2$ optimizing the superposition (green dashed line) can raise the entanglement entropy to be almost the same as for equal particle numbers (solid purple line with markers). (c) Degeneracies in the spectrum that mark crossing of spectral lines that include the ground state, coincide with discontinuous jumps in the entropy, shown here for an example in panel (b).}
\label{Fig4_chi}
\end{figure*}

The density matrix, $\rho$ of the composite system is defined in this basis. We measure the degree of entanglement between the two species by computing the von Neumann entanglement entropy \cite{Bennett_entanglement} using the reduced density matrices $\rho_2={\rm Tr}_1(\rho)$ or $\rho_1={\rm Tr}_2(\rho)$
\bn S(\rho_2)=-Tr[\rho_2\ln(\rho_2)]=-\sum_i[\epsilon_i\ln(\epsilon_i)]\en
The last step follows from assuming the density matrix can be diagonalized and ${\epsilon_i}$ are its eigenvalues. The entropy is not sensitive to the choice of the reduced density matrix $S(\rho_1)=S(\rho_2)$.

We compute the variation of the entropy with respect to the imbalance of the particle number and present them in Fig.~\ref{Fig3_imbalance}. This underscores another advantage of a system of equal number of particles in both species.  The entropy is maximized when $N_1=N_2$, as shown for two separate values of $N_1$ fixed as $N_2$ is varied. The maximum entropy is set by the smaller particle number $S_{max}=\ln|\min(N1,N2)|$.  The entanglement entropy is computed analytically form the solution that appears in Eq.~(\ref{exact_gs}) in the next section. An inherent degeneracy present in the ground state for unequal particle number underestimates the entropy for any specific ground state. We correct for this by adding $\ln|2|$ to allow for the degeneracy. When the imbalance is high, we find this match almost exactly the numerically computed entropy that optimizes for the linear combination of the degenerate ground states, suggesting equal weights maximizes the entropy.  However, close to equal number of particles, addition of $\ln|2|$ generally overestimates the entropy and the optimal entropy is not necessarily and equal weight combination the degenerate analytical solutions.

In Fig.~\ref{Fig4_chi}, we probe the sensitivity to our assumption equal interaction strengths, by plotting the entanglement entropy as we vary one of $\chi_\alpha$ keeping the other two fixed.  When we vary $\chi_2$ keeping $\chi_1$ and $\chi_{12}$ fixed, for both equal and unequal number of atoms, we find as seen in panel (a) the entropy decreases faster when $\chi_2$ is larger. On the other hand when we vary $\chi_{12}$ with other two fixed, panel (b) shows that the entropy drops off faster when $\chi_{12}$ is larger.  Therefore we can conclude that if there is a difference in the interaction strengths, it is better to have the inter-species interaction to be stronger than the intra-species ones.  The numerical computation of the entropy occasionally displays discontinuous jumps. We illustrate in Fig.~\ref{Fig4_chi}(c) that those jumps correspond to degeneracies where the ground state changes identity due to different spectral lines crossing.

\begin{figure*}[t]
\centering
\includegraphics[width=\textwidth]{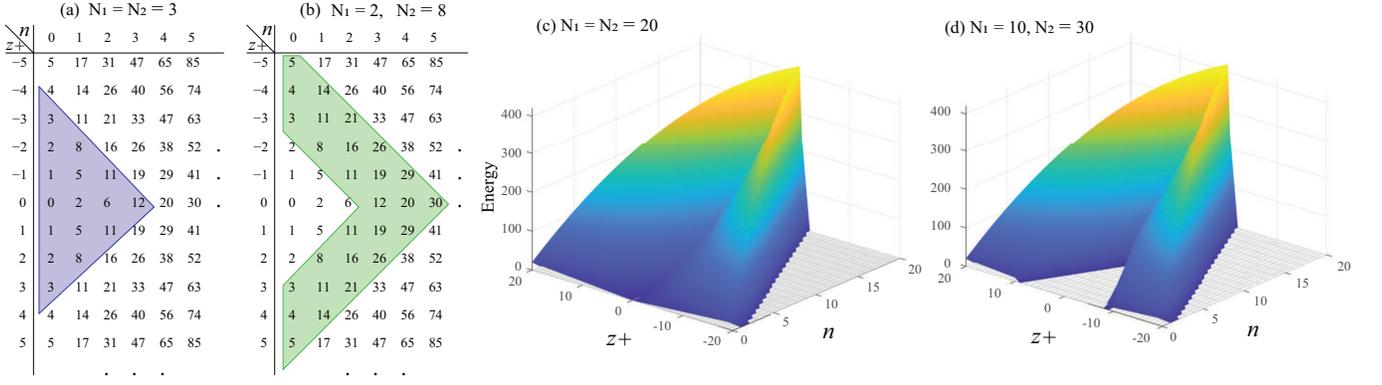}
\caption{(Color online) The eigenvalues of $\op{H}_Q$ are shown in the space of $z_+$ and $n$.  The allowed eigenvalues are shown by colored shading. (a) For $N_1=N_2$, for any allowed $z_+$ the minimum value of $z_-$ is always $0$. (b) For $N_1\neq N_2$, there is a regime of the lower $|z_+|$ where the lowest natural numbers including zero are excluded, creating a boomerang shape. The energy is plotted as a function of $z+$ and $n$ for (c) equal particle number $N_1=N_2$, when there is no gap at $n=0$ and (d) for unequal particle number $N_1\neq N_2$ when a gap emerges for lower $z+$ values.}
\label{Fig5_quantum_numbers}
\end{figure*}

\section{Analytical Eigenvalues and States}
In the case of all the couplings being the same, the quadratic Hamiltonian $\op{H}_Q$ in Eq.~(\ref{special_Hamiltonian}) can be diagonalized exactly. In the basis $|z_+,z_-\rangle$ defined above the Hamiltonian acquires a block tridiagonal structure
\begin{eqnarray}
\op{H}_Q |z_+,z_-\rangle &=& \left(n_{a1}n_{b1}+n_{a2}n_{b2}-{\textstyle \frac{1}{2}}N \right)|z_+,z_-\rangle
 \\
& +&
\sqrt{n_{a1}(n_{b1}+2) (n_{a2}+2)n_{b2}}|z_+,z_- - 2\rangle
\nonumber \\
&+&
\sqrt{(n_{a1}+2)n_{b1} n_{a2}(n_{b2}+2)}|z_+,z_- + 2\rangle ,
\nonumber\end{eqnarray}
where the $n_i$ are given by Eq.~(\ref{numbers}), and we define the total particle number $N=N_1+N_2$. Each block of fixed $z_+$ has triadiagonal form comprising set by the allowed $z_-$ values. We determine the eigenvalues to be given by
\begin{eqnarray}
E_n &=& n(n+1) + |z_+|(2n+1)\\
z_+ &=& 0, \pm 1, \pm 2, \dots \pm {\textstyle \frac{1}{2}}N\h{3mm}{\rm even}\ N\n\\
z_+ &=& {\textstyle \pm \frac{1}{2}, \pm \frac{3}{2}}, \dots \pm {\textstyle \frac{1}{2}}N \h{3mm}{\rm odd}\ N\n
\end{eqnarray}
where $n \in \{n_{\rm min},n_{\rm min}+2, \cdots,  n_{\rm max}\}$, with
\begin{eqnarray}
n_{\rm min} &=& \max \left( {\textstyle \frac{1}{2}}|N_2-N_1|- |z_+|, 0 \right), \n\\
n_{\rm max} &=& {\textstyle \frac{1}{2}}N - |z_+|.
\end{eqnarray}
This confirms explicitly some of the conclusions of the numerical results displayed in Fig.~\ref{Fig2_linear_to_quad}: When $N_1=N_2$, the expressions above shows that the ground state is indeed unique corresponding to $z_+=0$,
$n=0$ and energy $E_0=0$.  But, when $N_1 \neq N_2$, the lowest energy state is doubly degenerate, corresponding to $n=0$, but with
\begin{eqnarray}
z_+= \pm \frac{N_1-N_2}{2}, \qquad E_0 = \frac{|N_1-N_2|}{2}.
\end{eqnarray}
The eigenvalues depend on the atomic numbers $N_{1,2}$ only through the limits for the index $n$, as illustrated in Fig.~\ref{Fig5_quantum_numbers}. Since all the eigenvalues are integers or semi-integers with their smallest nonzero difference being 1, the evolution of any state is periodic with  period $2\pi$, assuring periodic behavior. This contrasts with a semiclassical description that will be reported in an upcoming work which suggests that the period should go to infinity.

Without loss of generality, we assume $N_1 \leq N_2$, the ground state for arbitrary particle numbers for the two species can be expressed in terms of the basis states $|z_+,z_-\rangle$ as
\begin{eqnarray}
|\psi_{0,\pm}\rangle &=& \sum_{k=0}^{N_1}\alpha_k \left|\pm {\textstyle \frac{1}{2}}(N_2-N_1),\mp \left[  {\textstyle \frac{1}{2}}(N_2-3N_1) + 2k \right] \right\rangle\n\\
\frac{\alpha_k}{\alpha_{k-1}} &=& -\sqrt{\frac{N_2-N_1+k}{k}},
\label{exact_gs}\end{eqnarray}
where the coefficients $\alpha_k$ are defined recursively. This formula also covers the special case $N_1=N_2=\frac{1}{2}N$, when the ground state becomes nondegenerate, with energy $E_0=0$ and $z_+=0$. The expressions then reduce to a simpler from which can be written as a superposition of states $|z_-\rangle$
\begin{eqnarray}
|\psi_0\rangle = \frac{\sqrt{2}}{\sqrt{N}}\sum_{k=0}^{\frac{1}{2}N} (-1)^k \left| -{\textstyle \frac{1}{2}}N+2k \right\rangle .
\label{EGroundN1eqN2}\end{eqnarray}

Beyond the ground state, in the special case of equal number of particles, $N_1=N_2=\frac{1}{2}N$ and in the subspace of $z_+=0$, which means there are equal number of counter-propagating atoms as well,  the energy is simply $E_n = n(n+1)$ and all the complete set of states in the subspace  are given by \bn |z_-\rangle={\textstyle |-\frac{1}{2}N+2n\rangle, \quad
n\in\{0,1,\cdots\frac{1}{2}N\}}.\en
This has an interesting implication for the dynamics. Since now all the eigenvalues are even integers and the minimum energy difference is 2, the evolution of any state is periodic with half the period $\pi$ compared to the more general case above. 

We conclude this section by noting that for \emph{minimal} asymmetry, $N_2=N_1+1$ the ground states have energy $E_0=\frac{1}{2}$ and correspond to $z_+=\pm \frac{1}{2}$. Expressed as superpositions of states $|z_+, z_-\rangle$ they are
\begin{eqnarray}
|\psi_{0,\pm}\rangle & =& \frac{\sqrt{2}}{\sqrt{N(N+1)}}
 \\ \nonumber
&& \times
\sum_{k=0}^{N_1}(-1)^k
\sqrt{k+1}\left| \pm {\textstyle \frac{1}{2}}, \pm \left( N_1-{\textstyle \frac{1}{2}} - 2k \right) \right\rangle .
\end{eqnarray}

\section{Limiting cases}

We now underscore the broad relevance of this Hamiltonian by identifying some limiting cases for the quadratic part $\op{H}_Q$. For this purpose, it is more transparent to express it in terms of the creation and annihilation operators
\begin{eqnarray}
\op{H}_Q&=&  \op{a}_1^{\dag}\op{a}_1 \op{b}_1^{\dag}\op{b}_1 + \op{a}_2^{\dag}\op{a}_2 \op{b}_2^{\dag}\op{b}_2 + {\textstyle\frac{1}{2}}(N_1+N_2)
\nonumber \\
&& + \op{a}_1 \op{b}_1^{\dag}\op{a}_2^{\dag}\op{b}_2 + \op{a}_1^{\dag}\op{b}_1 \op{a}_2 \op{b}_2^{\dag}.
\label{HamABdag}
\end{eqnarray}

\emph{Beam splitter limit}: If almost all the atoms in both species are circulating in the same direction, such that $b$-modes, $b_1\approx  b_1^{\dag} \approx \sqrt{N_1}$,  $b_2\approx  b_2^{\dag} \approx \sqrt{N_2}$, then the Hamiltonian reduces to
\begin{eqnarray}
\op{H}_Q&\approx&  N_1 \op{a}_1^{\dag}\op{a}_1  + N_2 \op{a}_2^{\dag}\op{a}_2 + {\textstyle\frac{1}{2}}(N_1+N_2)
\nonumber \\
&& + \sqrt{N_1 N_2} (\op{a}_1 \op{a}_2^{\dag} + \op{a}_1^{\dag}\op{a}_2 ).
\end{eqnarray}
The last term corresponds to a beam splitter (or linear coupler) which destroys one quantum (photon, for optical implementation) in one mode while creating one quantum in another mode (for details of the transformation, see, e.g., \cite{leonhardt_2010}). The first two terms are responsible for the time dependent change of phase in the two modes, the prefactors $N_{1,2}$ playing the role of frequencies of the modes. For $N_1=N_2\equiv N$ (matched frequencies) the Hamiltonian leads to oscillations of the mode occupations with period $\pi/N$ so that for time equal to $\pi/(2N)$ the atomic states are exchanged and for time equal to $\pi/(4N)$ the transformation corresponds to a 50/50 beam splitter
which can be used as a component to implement a Mach-Zehnder interferometer.  In Bloch sphere representation, the two species would be both lined towards the same pole.

\emph{Two-mode squeezer limit}: If almost all the atoms in the two species are circulating in opposite directions modes $\op{b}_1\simeq  \op{b}_1^{\dag}\simeq\sqrt{N_1}$ and $\op{a}_2\simeq  \op{a}_2^{\dag}\simeq\sqrt{N_2}$ (in Bloch sphere representation, the two species would be both lined towards opposite poles), we have
\begin{eqnarray}
\op{H}_Q&\approx&  N_1 \op{a}_1^{\dag}\op{a}_1  + N_2 \op{b}_2^{\dag}\op{b}_2 + {\textstyle\frac{1}{2}}(N_1+N_2)
\nonumber \\
&& + \sqrt{N_1 N_2} (\op{a}_1 \op{b}_2 + \op{a}_1^{\dag}\op{b}_2^{\dag} ).
\label{H2sqeez}
\end{eqnarray}
Here the last term crates or destroys pairs of quanta
in analogy to a parametric amplifier or a two-mode squeezer \cite{leonhardt_2010}.
This element could be used, e.g., to create highly entangled states of the atomic samples which metrological applications. If one can vary the sign of the nonlinearity, one can build a SU(1,1) interferometer \cite{Klauder} as a sequence of steps where first a squeezing Hamiltonian is applied, then a phase shifter (the phase of which is to be determined), and finally an un-squeezing Hamiltonian, which will require the opposite sign of the nonlinearity $\chi_{12}$.

\begin{figure}[t]
\centering
\includegraphics[width=\columnwidth]{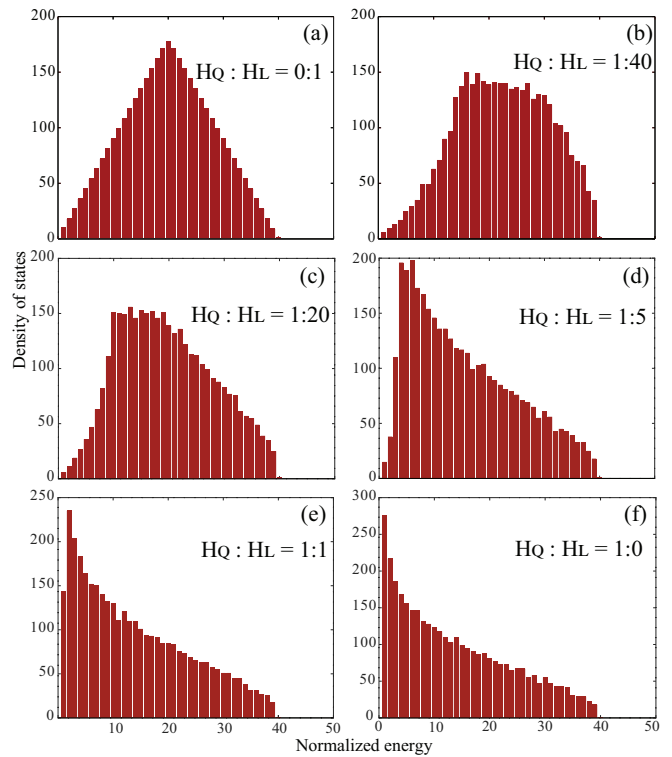}
\caption{(Color online) The distribution of energies, is shown for the case of $N_1 = N_2 = 59$, as we vary the full Hamiltonian in Eq.~(\ref{special_Hamiltonian}) from being purely linear, $\hat{H}=\hat{H_L}$ in panel (a) to being purely quadratic  $\hat{H}=\hat{H}_Q$ in panel (f). In the linear limit, we assume $\op{H}_L= (\op{J}_{z1}-\op{J}_{z2})$ in Eq.~(\ref{special_Hamiltonian}), corresponding to a rotating ring with no azimuthal lattice. }\label{Fig6_density}
\end{figure}

\section{Density of states}

While the variation of the spectrum in ranging from the linear to the quadratic Hamiltonian showed the degeneracy structure of the ground state, other significant differences can be identified by examining the density of states.  In Fig.~\ref{Fig6_density}, we plot the distribution of the energies as we adjust from  purely linear to the purely quadratic Hamiltonian. There is a marked difference. In the linear limit, the distribution shows a peak in the middle of the spectrum stemming from the fact that the energy eigenstates are the Dicke states of the two species with flat energy spectra. Combining these two individual spectra yields the largest number of possibilities for the middle value of the energy.
In the purely quadratic limit, the distribution is strongly skewed towards the ground state.  This follows from the energy function as shown in Fig.~\ref{Fig5_quantum_numbers} where large areas of parameters $z_+$ and $n$ correspond to small energy values. There is a gradual morphing of the distribution as we transition from one limit to the other. The disappearance of the peak and occurrence of a monotonously decreasing spectrum is suggestive of an excited state  quantum phase transition in the system \cite{Cejnar_2021}.

\section{Conclusions and Outlook}
Most of the analytical results we derived assume the interatomic interactions strengths to be equal $\chi_{1}=\chi_{2}=\chi_{12}$.  This is a reasonable assumption: For example, consider the hyperfine states $|F,m_F\rangle =|1, -1\rangle$ and  $|F,m_F\rangle =|2, 1\rangle$ as the two species, then all the scattering lengths are close to $a=100\ a_0$ \cite{Egorov_Rb}. Assuming a ring of major radius $R\sim 10\ \mu$m
and transverse trap frequency of $\omega=2\pi\times 100$ Hz, such as used in some recent experiments \cite{ramanathan} yields an interaction strength of $\chi=a\omega/(\pi R)\simeq 0.2$ Hz a value in the range used in our simulations. Of course, current technology allows for all of the parameters to be adjusted substantially, but this underscores the general experimental feasibility of our results.

Our analysis here shows that two species of ultracold atoms in a ring trap can provide a viable alternate platform to examine non-trivial quantum features that rely on entanglement. Here we mapped out the static and spectral properties as a necessary preliminary to examining the dynamical phenomena that can exploit the entanglement, which we are actively exploring in our continuing work. Among such applications, we already identified here certain limiting cases that can be adapted for interferometry as well as for generating two-mode squeezing. We are also currently examining ways to utilize the entangled states in this system to implement quantum teleportation \cite{Wootters_teleportation}, particularly the continuous variable variant in the limit of larger number of atoms \cite{Braunstein_Kimble}. With regards to all such quantum phenomena involving entangled states, the ring system offers the opportunity to study them in the context of external states encapsulated in circulating modes in the ring, rather than with internal states like spin typically utilized in the majority of platforms studied. This can facilitate a natural scaling up of the system size and the time scales involved, that can help better understand some of the most intriguing aspects of quantum mechanics.

\begin{acknowledgments} This work was supported by the Czech Science
Foundation Grant No. 20-27994S for T. Opatrn\'y and by the NSF under Grant No. PHY-2011767 for Kunal K. Das. \end{acknowledgments}


\begin{thebibliography}{42}%
\makeatletter
\providecommand \@ifxundefined [1]{%
 \@ifx{#1\undefined}
}%
\providecommand \@ifnum [1]{%
 \ifnum #1\expandafter \@firstoftwo
 \else \expandafter \@secondoftwo
 \fi
}%
\providecommand \@ifx [1]{%
 \ifx #1\expandafter \@firstoftwo
 \else \expandafter \@secondoftwo
 \fi
}%
\providecommand \natexlab [1]{#1}%
\providecommand \enquote  [1]{``#1''}%
\providecommand \bibnamefont  [1]{#1}%
\providecommand \bibfnamefont [1]{#1}%
\providecommand \citenamefont [1]{#1}%
\providecommand \href@noop [0]{\@secondoftwo}%
\providecommand \href [0]{\begingroup \@sanitize@url \@href}%
\providecommand \@href[1]{\@@startlink{#1}\@@href}%
\providecommand \@@href[1]{\endgroup#1\@@endlink}%
\providecommand \@sanitize@url [0]{\catcode `\\12\catcode `\$12\catcode
  `\&12\catcode `\#12\catcode `\^12\catcode `\_12\catcode `\%12\relax}%
\providecommand \@@startlink[1]{}%
\providecommand \@@endlink[0]{}%
\providecommand \url  [0]{\begingroup\@sanitize@url \@url }%
\providecommand \@url [1]{\endgroup\@href {#1}{\urlprefix }}%
\providecommand \urlprefix  [0]{URL }%
\providecommand \Eprint [0]{\href }%
\providecommand \doibase [0]{https://doi.org/}%
\providecommand \selectlanguage [0]{\@gobble}%
\providecommand \bibinfo  [0]{\@secondoftwo}%
\providecommand \bibfield  [0]{\@secondoftwo}%
\providecommand \translation [1]{[#1]}%
\providecommand \BibitemOpen [0]{}%
\providecommand \bibitemStop [0]{}%
\providecommand \bibitemNoStop [0]{.\EOS\space}%
\providecommand \EOS [0]{\spacefactor3000\relax}%
\providecommand \BibitemShut  [1]{\csname bibitem#1\endcsname}%
\let\auto@bib@innerbib\@empty
\bibitem [{\citenamefont {Ramanathan}\ \emph {et~al.}(2011)\citenamefont
  {Ramanathan}, \citenamefont {Wright}, \citenamefont {Muniz}, \citenamefont
  {Zelan}, \citenamefont {Hill}, \citenamefont {Lobb}, \citenamefont
  {Helmerson}, \citenamefont {Phillips},\ and\ \citenamefont
  {Campbell}}]{ramanathan}%
  \BibitemOpen
  \bibfield  {author} {\bibinfo {author} {\bibfnamefont {A.}~\bibnamefont
  {Ramanathan}}, \bibinfo {author} {\bibfnamefont {K.~C.}\ \bibnamefont
  {Wright}}, \bibinfo {author} {\bibfnamefont {S.~R.}\ \bibnamefont {Muniz}},
  \bibinfo {author} {\bibfnamefont {M.}~\bibnamefont {Zelan}}, \bibinfo
  {author} {\bibfnamefont {W.~T.}\ \bibnamefont {Hill}}, \bibinfo {author}
  {\bibfnamefont {C.~J.}\ \bibnamefont {Lobb}}, \bibinfo {author}
  {\bibfnamefont {K.}~\bibnamefont {Helmerson}}, \bibinfo {author}
  {\bibfnamefont {W.~D.}\ \bibnamefont {Phillips}},\ and\ \bibinfo {author}
  {\bibfnamefont {G.~K.}\ \bibnamefont {Campbell}},\ }\bibfield  {title}
  {\bibinfo {title} {Superflow in a toroidal Bose-Einstein condensate: An atom
  circuit with a tunable weak link},\ }\href
  {https://doi.org/10.1103/PhysRevLett.106.130401} {\bibfield  {journal}
  {\bibinfo  {journal} {Phys. Rev. Lett.}\ }\textbf {\bibinfo {volume} {106}},\
  \bibinfo {pages} {130401} (\bibinfo {year} {2011})}\BibitemShut {NoStop}%
\bibitem [{\citenamefont {Wright}\ \emph {et~al.}(2013)\citenamefont {Wright},
  \citenamefont {Blakestad}, \citenamefont {Lobb}, \citenamefont {Phillips},\
  and\ \citenamefont {Campbell}}]{Phillips_Campbell_superfluid_2013}%
  \BibitemOpen
  \bibfield  {author} {\bibinfo {author} {\bibfnamefont {K.~C.}\ \bibnamefont
  {Wright}}, \bibinfo {author} {\bibfnamefont {R.~B.}\ \bibnamefont
  {Blakestad}}, \bibinfo {author} {\bibfnamefont {C.~J.}\ \bibnamefont {Lobb}},
  \bibinfo {author} {\bibfnamefont {W.~D.}\ \bibnamefont {Phillips}},\ and\
  \bibinfo {author} {\bibfnamefont {G.~K.}\ \bibnamefont {Campbell}},\
  }\bibfield  {title} {\bibinfo {title} {Driving phase slips in a superfluid
  atom circuit with a rotating weak link},\ }\href
  {https://doi.org/10.1103/PhysRevLett.110.025302} {\bibfield  {journal}
  {\bibinfo  {journal} {Phys. Rev. Lett.}\ }\textbf {\bibinfo {volume} {110}},\
  \bibinfo {pages} {025302} (\bibinfo {year} {2013})}\BibitemShut {NoStop}%
\bibitem [{\citenamefont {Brooks}\ \emph {et~al.}(2021)\citenamefont {Brooks},
  \citenamefont {Brattley},\ and\ \citenamefont {Das}}]{Das-Brooks-Brattley}%
  \BibitemOpen
  \bibfield  {author} {\bibinfo {author} {\bibfnamefont {C.}~\bibnamefont
  {Brooks}}, \bibinfo {author} {\bibfnamefont {A.}~\bibnamefont {Brattley}},\
  and\ \bibinfo {author} {\bibfnamefont {K.~K.}\ \bibnamefont {Das}},\
  }\bibfield  {title} {\bibinfo {title} {Rotation-sensitive quench and revival
  of coherent oscillations in a ring lattice},\ }\href
  {https://doi.org/10.1103/PhysRevA.103.013322} {\bibfield  {journal} {\bibinfo
   {journal} {Phys. Rev. A}\ }\textbf {\bibinfo {volume} {103}},\ \bibinfo
  {pages} {013322} (\bibinfo {year} {2021})}\BibitemShut {NoStop}%
\bibitem [{\citenamefont {Bloch}\ \emph {et~al.}(2008)\citenamefont {Bloch},
  \citenamefont {Dalibard},\ and\ \citenamefont
  {Zwerger}}]{Bloch-RMP-Many-Body}%
  \BibitemOpen
  \bibfield  {author} {\bibinfo {author} {\bibfnamefont {I.}~\bibnamefont
  {Bloch}}, \bibinfo {author} {\bibfnamefont {J.}~\bibnamefont {Dalibard}},\
  and\ \bibinfo {author} {\bibfnamefont {W.}~\bibnamefont {Zwerger}},\
  }\bibfield  {title} {\bibinfo {title} {Many-body physics with ultracold
  gases},\ }\href {https://doi.org/10.1103/RevModPhys.80.885} {\bibfield
  {journal} {\bibinfo  {journal} {Rev. Mod. Phys.}\ }\textbf {\bibinfo {volume}
  {80}},\ \bibinfo {pages} {885} (\bibinfo {year} {2008})}\BibitemShut
  {NoStop}%
\bibitem [{\citenamefont {Huang}\ and\ \citenamefont {Das}(2021)}]{Das-Huang}%
  \BibitemOpen
  \bibfield  {author} {\bibinfo {author} {\bibfnamefont {H.}~\bibnamefont
  {Huang}}\ and\ \bibinfo {author} {\bibfnamefont {K.~K.}\ \bibnamefont
  {Das}},\ }\bibfield  {title} {\bibinfo {title} {Effects of a rotating
  periodic lattice on coherent quantum states in a ring topology: The case of
  positive nonlinearity},\ }\href {https://doi.org/10.1103/PhysRevA.104.053320}
  {\bibfield  {journal} {\bibinfo  {journal} {Phys. Rev. A}\ }\textbf {\bibinfo
  {volume} {104}},\ \bibinfo {pages} {053320} (\bibinfo {year}
  {2021})}\BibitemShut {NoStop}%
\bibitem [{\citenamefont {Kitagawa}\ and\ \citenamefont
  {Ueda}(1993)}]{Kitagawa}%
  \BibitemOpen
  \bibfield  {author} {\bibinfo {author} {\bibfnamefont {M.}~\bibnamefont
  {Kitagawa}}\ and\ \bibinfo {author} {\bibfnamefont {M.}~\bibnamefont
  {Ueda}},\ }\bibfield  {title} {\bibinfo {title} {Squeezed spin states},\
  }\href {https://doi.org/10.1103/PhysRevA.47.5138} {\bibfield  {journal}
  {\bibinfo  {journal} {Phys. Rev. A}\ }\textbf {\bibinfo {volume} {47}},\
  \bibinfo {pages} {5138} (\bibinfo {year} {1993})}\BibitemShut {NoStop}%
\bibitem [{\citenamefont {Gupta}\ \emph {et~al.}(2005)\citenamefont {Gupta},
  \citenamefont {Murch}, \citenamefont {Moore}, \citenamefont {Purdy},\ and\
  \citenamefont {Stamper-Kurn}}]{Stamper-Kurn-2005}%
  \BibitemOpen
  \bibfield  {author} {\bibinfo {author} {\bibfnamefont {S.}~\bibnamefont
  {Gupta}}, \bibinfo {author} {\bibfnamefont {K.~W.}\ \bibnamefont {Murch}},
  \bibinfo {author} {\bibfnamefont {K.~L.}\ \bibnamefont {Moore}}, \bibinfo
  {author} {\bibfnamefont {T.~P.}\ \bibnamefont {Purdy}},\ and\ \bibinfo
  {author} {\bibfnamefont {D.~M.}\ \bibnamefont {Stamper-Kurn}},\ }\bibfield
  {title} {\bibinfo {title} {Bose-Einstein condensation in a circular
  waveguide},\ }\href {https://doi.org/10.1103/PhysRevLett.95.143201}
  {\bibfield  {journal} {\bibinfo  {journal} {Phys. Rev. Lett.}\ }\textbf
  {\bibinfo {volume} {95}},\ \bibinfo {pages} {143201} (\bibinfo {year}
  {2005})}\BibitemShut {NoStop}%
\bibitem [{\citenamefont {Arnold}\ \emph {et~al.}(2006)\citenamefont {Arnold},
  \citenamefont {Garvie},\ and\ \citenamefont {Riis}}]{Riis-magnetic}%
  \BibitemOpen
  \bibfield  {author} {\bibinfo {author} {\bibfnamefont {A.~S.}\ \bibnamefont
  {Arnold}}, \bibinfo {author} {\bibfnamefont {C.~S.}\ \bibnamefont {Garvie}},\
  and\ \bibinfo {author} {\bibfnamefont {E.}~\bibnamefont {Riis}},\ }\bibfield
  {title} {\bibinfo {title} {Large magnetic storage ring for Bose-Einstein
  condensates},\ }\href {https://doi.org/10.1103/PhysRevA.73.041606} {\bibfield
   {journal} {\bibinfo  {journal} {Phys. Rev. A}\ }\textbf {\bibinfo {volume}
  {73}},\ \bibinfo {pages} {041606(R)} (\bibinfo {year} {2006})}\BibitemShut
  {NoStop}%
\bibitem [{\citenamefont {Sauer}\ \emph {et~al.}(2001)\citenamefont {Sauer},
  \citenamefont {Barrett},\ and\ \citenamefont
  {Chapman}}]{Chapman-magnetic-ring-2001}%
  \BibitemOpen
  \bibfield  {author} {\bibinfo {author} {\bibfnamefont {J.~A.}\ \bibnamefont
  {Sauer}}, \bibinfo {author} {\bibfnamefont {M.~D.}\ \bibnamefont {Barrett}},\
  and\ \bibinfo {author} {\bibfnamefont {M.~S.}\ \bibnamefont {Chapman}},\
  }\bibfield  {title} {\bibinfo {title} {Storage ring for neutral atoms},\
  }\href {https://doi.org/10.1103/PhysRevLett.87.270401} {\bibfield  {journal}
  {\bibinfo  {journal} {Phys. Rev. Lett.}\ }\textbf {\bibinfo {volume} {87}},\
  \bibinfo {pages} {270401} (\bibinfo {year} {2001})}\BibitemShut {NoStop}%
\bibitem [{\citenamefont {Henderson}\ \emph {et~al.}(2009)\citenamefont
  {Henderson}, \citenamefont {Ryu}, \citenamefont {MacCormick},\ and\
  \citenamefont {Boshier}}]{Boshier-painted-potential}%
  \BibitemOpen
  \bibfield  {author} {\bibinfo {author} {\bibfnamefont {K.}~\bibnamefont
  {Henderson}}, \bibinfo {author} {\bibfnamefont {C.}~\bibnamefont {Ryu}},
  \bibinfo {author} {\bibfnamefont {C.}~\bibnamefont {MacCormick}},\ and\
  \bibinfo {author} {\bibfnamefont {M.~G.}\ \bibnamefont {Boshier}},\
  }\bibfield  {title} {\bibinfo {title} {Experimental demonstration of painting
  arbitrary and dynamic potentials for Bose–Einstein condensates},\
  }\href@noop {} {\bibfield  {journal} {\bibinfo  {journal} {New Journal of
  Physics}\ }\textbf {\bibinfo {volume} {11}},\ \bibinfo {pages} {043030}
  (\bibinfo {year} {2009})}\BibitemShut {NoStop}%
\bibitem [{\citenamefont {Sherlock}\ \emph {et~al.}(2011)\citenamefont
  {Sherlock}, \citenamefont {Gildemeister}, \citenamefont {Owen}, \citenamefont
  {Nugent},\ and\ \citenamefont {Foot}}]{Foot-painted}%
  \BibitemOpen
  \bibfield  {author} {\bibinfo {author} {\bibfnamefont {B.~E.}\ \bibnamefont
  {Sherlock}}, \bibinfo {author} {\bibfnamefont {M.}~\bibnamefont
  {Gildemeister}}, \bibinfo {author} {\bibfnamefont {E.}~\bibnamefont {Owen}},
  \bibinfo {author} {\bibfnamefont {E.}~\bibnamefont {Nugent}},\ and\ \bibinfo
  {author} {\bibfnamefont {C.~J.}\ \bibnamefont {Foot}},\ }\bibfield  {title}
  {\bibinfo {title} {Time-averaged adiabatic ring potential for ultracold
  atoms},\ }\href {https://doi.org/10.1103/PhysRevA.83.043408} {\bibfield
  {journal} {\bibinfo  {journal} {Phys. Rev. A}\ }\textbf {\bibinfo {volume}
  {83}},\ \bibinfo {pages} {043408} (\bibinfo {year} {2011})}\BibitemShut
  {NoStop}%
\bibitem [{\citenamefont {Lee}\ and\ \citenamefont
  {Hill}(2014)}]{Lee_Hill_photomask}%
  \BibitemOpen
  \bibfield  {author} {\bibinfo {author} {\bibfnamefont {J.~G.}\ \bibnamefont
  {Lee}}\ and\ \bibinfo {author} {\bibfnamefont {W.~T.}\ \bibnamefont {Hill}},\
  }\bibfield  {title} {\bibinfo {title} {Spatial shaping for generating
  arbitrary optical dipole traps for ultracold degenerate gases},\ }\href@noop
  {} {\bibfield  {journal} {\bibinfo  {journal} {Review of Scientific
  Instruments}\ }\textbf {\bibinfo {volume} {85}},\ \bibinfo {eid} {103106}
  (\bibinfo {year} {2014})}\BibitemShut {NoStop}%
\bibitem [{\citenamefont {Turpin}\ \emph {et~al.}(2015)\citenamefont {Turpin},
  \citenamefont {Polo}, \citenamefont {Loiko}, \citenamefont {K\"{u}ber},
  \citenamefont {Schmaltz}, \citenamefont {Kalkandjiev}, \citenamefont
  {Ahufinger}, \citenamefont {Birkl},\ and\ \citenamefont
  {Mompart}}]{Mompart-conical}%
  \BibitemOpen
  \bibfield  {author} {\bibinfo {author} {\bibfnamefont {A.}~\bibnamefont
  {Turpin}}, \bibinfo {author} {\bibfnamefont {J.}~\bibnamefont {Polo}},
  \bibinfo {author} {\bibfnamefont {Y.~V.}\ \bibnamefont {Loiko}}, \bibinfo
  {author} {\bibfnamefont {J.}~\bibnamefont {K\"{u}ber}}, \bibinfo {author}
  {\bibfnamefont {F.}~\bibnamefont {Schmaltz}}, \bibinfo {author}
  {\bibfnamefont {T.~K.}\ \bibnamefont {Kalkandjiev}}, \bibinfo {author}
  {\bibfnamefont {V.}~\bibnamefont {Ahufinger}}, \bibinfo {author}
  {\bibfnamefont {G.}~\bibnamefont {Birkl}},\ and\ \bibinfo {author}
  {\bibfnamefont {J.}~\bibnamefont {Mompart}},\ }\bibfield  {title} {\bibinfo
  {title} {Blue-detuned optical ring trap for Bose-Einstein condensates based
  on conical refraction},\ }\href@noop {} {\bibfield  {journal} {\bibinfo
  {journal} {Opt. Express}\ }\textbf {\bibinfo {volume} {23}},\ \bibinfo
  {pages} {1638} (\bibinfo {year} {2015})}\BibitemShut {NoStop}%
\bibitem [{\citenamefont {Bell}\ \emph {et~al.}(2016)\citenamefont {Bell},
  \citenamefont {Glidden}, \citenamefont {Humbert}, \citenamefont {Bromley},
  \citenamefont {Haine}, \citenamefont {Davis}, \citenamefont {Neely},
  \citenamefont {Baker},\ and\ \citenamefont
  {Rubinsztein-Dunlop}}]{Dunlop-ring}%
  \BibitemOpen
  \bibfield  {author} {\bibinfo {author} {\bibfnamefont {T.~A.}\ \bibnamefont
  {Bell}}, \bibinfo {author} {\bibfnamefont {J.~A.~P.}\ \bibnamefont
  {Glidden}}, \bibinfo {author} {\bibfnamefont {L.}~\bibnamefont {Humbert}},
  \bibinfo {author} {\bibfnamefont {M.~W.~J.}\ \bibnamefont {Bromley}},
  \bibinfo {author} {\bibfnamefont {S.~A.}\ \bibnamefont {Haine}}, \bibinfo
  {author} {\bibfnamefont {M.~J.}\ \bibnamefont {Davis}}, \bibinfo {author}
  {\bibfnamefont {T.~W.}\ \bibnamefont {Neely}}, \bibinfo {author}
  {\bibfnamefont {M.~A.}\ \bibnamefont {Baker}},\ and\ \bibinfo {author}
  {\bibfnamefont {H.}~\bibnamefont {Rubinsztein-Dunlop}},\ }\bibfield  {title}
  {\bibinfo {title} {Bose-Einstein condensation in large time-averaged optical
  ring potentials},\ }\href@noop {} {\bibfield  {journal} {\bibinfo  {journal}
  {New Journal of Physics}\ }\textbf {\bibinfo {volume} {18}},\ \bibinfo
  {pages} {035003} (\bibinfo {year} {2016})}\BibitemShut {NoStop}%
\bibitem [{\citenamefont {Meister}\ \emph {et~al.}(2017)\citenamefont
  {Meister}, \citenamefont {Arnold}, \citenamefont {Moll}, \citenamefont
  {Eckart}, \citenamefont {Kajari}, \citenamefont {Efremov}, \citenamefont
  {Walser},\ and\ \citenamefont {Schleich}}]{MEISTER2017375}%
  \BibitemOpen
  \bibfield  {author} {\bibinfo {author} {\bibfnamefont {M.}~\bibnamefont
  {Meister}}, \bibinfo {author} {\bibfnamefont {S.}~\bibnamefont {Arnold}},
  \bibinfo {author} {\bibfnamefont {D.}~\bibnamefont {Moll}}, \bibinfo {author}
  {\bibfnamefont {M.}~\bibnamefont {Eckart}}, \bibinfo {author} {\bibfnamefont
  {E.}~\bibnamefont {Kajari}}, \bibinfo {author} {\bibfnamefont {M.~A.}\
  \bibnamefont {Efremov}}, \bibinfo {author} {\bibfnamefont {R.}~\bibnamefont
  {Walser}},\ and\ \bibinfo {author} {\bibfnamefont {W.~P.}\ \bibnamefont
  {Schleich}},\ }\href@noop {} {\emph {\bibinfo {title} {Efficient Description
  of Bose-Einstein Condensates in Time-Dependent Rotating Traps}}},\ edited by\
  \bibinfo {editor} {\bibfnamefont {E.}~\bibnamefont {Arimondo}}, \bibinfo
  {editor} {\bibfnamefont {C.~C.}\ \bibnamefont {Lin}},\ and\ \bibinfo {editor}
  {\bibfnamefont {S.~F.}\ \bibnamefont {Yelin}},\ \bibinfo {series} {Advances
  In Atomic, Molecular, and Optical Physics}, Vol.~\bibinfo {volume} {66}\
  (\bibinfo  {publisher} {Academic Press},\ \bibinfo {year} {2017})\ pp.\
  \bibinfo {pages} {375--438}\BibitemShut {NoStop}%
\bibitem [{\citenamefont {Franke-Arnold}\ \emph {et~al.}(2007)\citenamefont
  {Franke-Arnold}, \citenamefont {Leach}, \citenamefont {Padgett},
  \citenamefont {Lembessis}, \citenamefont {Ellinas}, \citenamefont {Wright},
  \citenamefont {Girkin}, \citenamefont {\"{O}hberg},\ and\ \citenamefont
  {Arnold}}]{Padgett}%
  \BibitemOpen
  \bibfield  {author} {\bibinfo {author} {\bibfnamefont {S.}~\bibnamefont
  {Franke-Arnold}}, \bibinfo {author} {\bibfnamefont {J.}~\bibnamefont
  {Leach}}, \bibinfo {author} {\bibfnamefont {M.~J.}\ \bibnamefont {Padgett}},
  \bibinfo {author} {\bibfnamefont {V.~E.}\ \bibnamefont {Lembessis}}, \bibinfo
  {author} {\bibfnamefont {D.}~\bibnamefont {Ellinas}}, \bibinfo {author}
  {\bibfnamefont {A.~J.}\ \bibnamefont {Wright}}, \bibinfo {author}
  {\bibfnamefont {J.~M.}\ \bibnamefont {Girkin}}, \bibinfo {author}
  {\bibfnamefont {P.}~\bibnamefont {\"{O}hberg}},\ and\ \bibinfo {author}
  {\bibfnamefont {A.~S.}\ \bibnamefont {Arnold}},\ }\bibfield  {title}
  {\bibinfo {title} {Optical Ferris wheel for ultracold atoms},\ }\href@noop {}
  {\bibfield  {journal} {\bibinfo  {journal} {Opt. Express}\ }\textbf {\bibinfo
  {volume} {15}},\ \bibinfo {pages} {8619} (\bibinfo {year}
  {2007})}\BibitemShut {NoStop}%
\bibitem [{\citenamefont {Zambrini}\ and\ \citenamefont
  {Barnett}(2007)}]{Zambrini:07}%
  \BibitemOpen
  \bibfield  {author} {\bibinfo {author} {\bibfnamefont {R.}~\bibnamefont
  {Zambrini}}\ and\ \bibinfo {author} {\bibfnamefont {S.~M.}\ \bibnamefont
  {Barnett}},\ }\bibfield  {title} {\bibinfo {title} {Angular momentum of
  multimode and polarization patterns},\ }\href@noop {} {\bibfield  {journal}
  {\bibinfo  {journal} {Opt. Express}\ }\textbf {\bibinfo {volume} {15}},\
  \bibinfo {pages} {15214} (\bibinfo {year} {2007})}\BibitemShut {NoStop}%
\bibitem [{\citenamefont {Jendrzejewski}\ \emph {et~al.}(2014)\citenamefont
  {Jendrzejewski}, \citenamefont {Eckel}, \citenamefont {Murray}, \citenamefont
  {Lanier}, \citenamefont {Edwards}, \citenamefont {Lobb},\ and\ \citenamefont
  {Campbell}}]{Campbell_resistive-flow}%
  \BibitemOpen
  \bibfield  {author} {\bibinfo {author} {\bibfnamefont {F.}~\bibnamefont
  {Jendrzejewski}}, \bibinfo {author} {\bibfnamefont {S.}~\bibnamefont
  {Eckel}}, \bibinfo {author} {\bibfnamefont {N.}~\bibnamefont {Murray}},
  \bibinfo {author} {\bibfnamefont {C.}~\bibnamefont {Lanier}}, \bibinfo
  {author} {\bibfnamefont {M.}~\bibnamefont {Edwards}}, \bibinfo {author}
  {\bibfnamefont {C.~J.}\ \bibnamefont {Lobb}},\ and\ \bibinfo {author}
  {\bibfnamefont {G.~K.}\ \bibnamefont {Campbell}},\ }\bibfield  {title}
  {\bibinfo {title} {Resistive flow in a weakly interacting Bose-Einstein
  condensate},\ }\href {https://doi.org/10.1103/PhysRevLett.113.045305}
  {\bibfield  {journal} {\bibinfo  {journal} {Phys. Rev. Lett.}\ }\textbf
  {\bibinfo {volume} {113}},\ \bibinfo {pages} {045305} (\bibinfo {year}
  {2014})}\BibitemShut {NoStop}%
\bibitem [{\citenamefont {Eckel}\ \emph {et~al.}(2014)\citenamefont {Eckel},
  \citenamefont {Lee}, \citenamefont {Jendrzejewski}, \citenamefont {Murray},
  \citenamefont {Clark}, \citenamefont {Lobb}, \citenamefont {Phillips},\ and\
  \citenamefont {Campbell}}]{Phillips_Campbell_hysteresis}%
  \BibitemOpen
  \bibfield  {author} {\bibinfo {author} {\bibfnamefont {S.}~\bibnamefont
  {Eckel}}, \bibinfo {author} {\bibfnamefont {J.~G.}\ \bibnamefont {Lee}},
  \bibinfo {author} {\bibfnamefont {F.}~\bibnamefont {Jendrzejewski}}, \bibinfo
  {author} {\bibfnamefont {N.}~\bibnamefont {Murray}}, \bibinfo {author}
  {\bibfnamefont {C.~W.}\ \bibnamefont {Clark}}, \bibinfo {author}
  {\bibfnamefont {C.~J.}\ \bibnamefont {Lobb}}, \bibinfo {author}
  {\bibfnamefont {M.}~\bibnamefont {Phillips}, \bibfnamefont
  {W.~D.and~Edwards}},\ and\ \bibinfo {author} {\bibfnamefont {G.~K.}\
  \bibnamefont {Campbell}},\ }\bibfield  {title} {\bibinfo {title} {Hysteresis
  in a quantized superfluid `atomtronic' circuit},\ }\href@noop {} {\bibfield
  {journal} {\bibinfo  {journal} {Nature}\ }\textbf {\bibinfo {volume} {506}},\
  \bibinfo {pages} {200} (\bibinfo {year} {2014})}\BibitemShut {NoStop}%
\bibitem [{\citenamefont {Aghamalyan}\ \emph {et~al.}(2015)\citenamefont
  {Aghamalyan}, \citenamefont {Cominotti}, \citenamefont {Rizzi}, \citenamefont
  {Rossini}, \citenamefont {Hekking}, \citenamefont {Minguzzi}, \citenamefont
  {Kwek},\ and\ \citenamefont {Amico}}]{Aghamalyan-AQUID}%
  \BibitemOpen
  \bibfield  {author} {\bibinfo {author} {\bibfnamefont {D.}~\bibnamefont
  {Aghamalyan}}, \bibinfo {author} {\bibfnamefont {M.}~\bibnamefont
  {Cominotti}}, \bibinfo {author} {\bibfnamefont {M.}~\bibnamefont {Rizzi}},
  \bibinfo {author} {\bibfnamefont {D.}~\bibnamefont {Rossini}}, \bibinfo
  {author} {\bibfnamefont {F.}~\bibnamefont {Hekking}}, \bibinfo {author}
  {\bibfnamefont {A.}~\bibnamefont {Minguzzi}}, \bibinfo {author}
  {\bibfnamefont {L.-C.}\ \bibnamefont {Kwek}},\ and\ \bibinfo {author}
  {\bibfnamefont {L.}~\bibnamefont {Amico}},\ }\bibfield  {title} {\bibinfo
  {title} {Coherent superposition of current flows in an atomtronic quantum
  interference device},\ }\href@noop {} {\bibfield  {journal} {\bibinfo
  {journal} {New Journal of Physics}\ }\textbf {\bibinfo {volume} {17}},\
  \bibinfo {pages} {045023} (\bibinfo {year} {2015})}\BibitemShut {NoStop}%
\bibitem [{\citenamefont {Aghamalyan}\ \emph {et~al.}(2013)\citenamefont
  {Aghamalyan}, \citenamefont {Amico},\ and\ \citenamefont
  {Kwek}}]{Aghamalyan-two-ring-lattice}%
  \BibitemOpen
  \bibfield  {author} {\bibinfo {author} {\bibfnamefont {D.}~\bibnamefont
  {Aghamalyan}}, \bibinfo {author} {\bibfnamefont {L.}~\bibnamefont {Amico}},\
  and\ \bibinfo {author} {\bibfnamefont {L.~C.}\ \bibnamefont {Kwek}},\
  }\bibfield  {title} {\bibinfo {title} {Effective dynamics of cold atoms
  flowing in two ring-shaped optical potentials with tunable tunneling},\
  }\href {https://doi.org/10.1103/PhysRevA.88.063627} {\bibfield  {journal}
  {\bibinfo  {journal} {Phys. Rev. A}\ }\textbf {\bibinfo {volume} {88}},\
  \bibinfo {pages} {063627} (\bibinfo {year} {2013})}\BibitemShut {NoStop}%
\bibitem [{\citenamefont {Pinheiro}\ and\ \citenamefont
  {de~Toledo~Piza}(2013)}]{Piza-ring-lattice}%
  \BibitemOpen
  \bibfield  {author} {\bibinfo {author} {\bibfnamefont {F.}~\bibnamefont
  {Pinheiro}}\ and\ \bibinfo {author} {\bibfnamefont {A.~F.~R.}\ \bibnamefont
  {de~Toledo~Piza}},\ }\bibfield  {title} {\bibinfo {title} {Delocalization and
  superfluidity of ultracold bosonic atoms in a ring lattice},\ }\href@noop {}
  {\bibfield  {journal} {\bibinfo  {journal} {Journal of Physics B: Atomic,
  Molecular and Optical Physics}\ }\textbf {\bibinfo {volume} {46}},\ \bibinfo
  {pages} {205303} (\bibinfo {year} {2013})}\BibitemShut {NoStop}%
\bibitem [{\citenamefont {Satija}\ \emph {et~al.}(2013)\citenamefont {Satija},
  \citenamefont {Pando~L.},\ and\ \citenamefont
  {Tiesinga}}]{Tiesinga-soliton-lattice}%
  \BibitemOpen
  \bibfield  {author} {\bibinfo {author} {\bibfnamefont {I.~I.}\ \bibnamefont
  {Satija}}, \bibinfo {author} {\bibfnamefont {C.~L.}\ \bibnamefont
  {Pando~L.}},\ and\ \bibinfo {author} {\bibfnamefont {E.}~\bibnamefont
  {Tiesinga}},\ }\bibfield  {title} {\bibinfo {title} {Soliton dynamics of an
  atomic spinor condensate on a ring lattice},\ }\href
  {https://doi.org/10.1103/PhysRevA.87.033608} {\bibfield  {journal} {\bibinfo
  {journal} {Phys. Rev. A}\ }\textbf {\bibinfo {volume} {87}},\ \bibinfo
  {pages} {033608} (\bibinfo {year} {2013})}\BibitemShut {NoStop}%
\bibitem [{\citenamefont {Maik}\ \emph {et~al.}(2011)\citenamefont {Maik},
  \citenamefont {Buonsante}, \citenamefont {Vezzani},\ and\ \citenamefont
  {Zakrzewski}}]{Maik-dipolar}%
  \BibitemOpen
  \bibfield  {author} {\bibinfo {author} {\bibfnamefont {M.}~\bibnamefont
  {Maik}}, \bibinfo {author} {\bibfnamefont {P.}~\bibnamefont {Buonsante}},
  \bibinfo {author} {\bibfnamefont {A.}~\bibnamefont {Vezzani}},\ and\ \bibinfo
  {author} {\bibfnamefont {J.}~\bibnamefont {Zakrzewski}},\ }\bibfield  {title}
  {\bibinfo {title} {Dipolar bosons on an optical lattice ring},\ }\href
  {https://doi.org/10.1103/PhysRevA.84.053615} {\bibfield  {journal} {\bibinfo
  {journal} {Phys. Rev. A}\ }\textbf {\bibinfo {volume} {84}},\ \bibinfo
  {pages} {053615} (\bibinfo {year} {2011})}\BibitemShut {NoStop}%
\bibitem [{\citenamefont {Hettiarachchilage}\ \emph {et~al.}(2013)\citenamefont
  {Hettiarachchilage}, \citenamefont {Rousseau}, \citenamefont {Tam},
  \citenamefont {Jarrell},\ and\ \citenamefont {Moreno}}]{Moreno-Bose-Hubbard}%
  \BibitemOpen
  \bibfield  {author} {\bibinfo {author} {\bibfnamefont {K.}~\bibnamefont
  {Hettiarachchilage}}, \bibinfo {author} {\bibfnamefont {V.~G.}\ \bibnamefont
  {Rousseau}}, \bibinfo {author} {\bibfnamefont {K.-M.}\ \bibnamefont {Tam}},
  \bibinfo {author} {\bibfnamefont {M.}~\bibnamefont {Jarrell}},\ and\ \bibinfo
  {author} {\bibfnamefont {J.}~\bibnamefont {Moreno}},\ }\bibfield  {title}
  {\bibinfo {title} {Phase diagram of the Bose-hubbard model on a ring-shaped
  lattice with tunable weak links},\ }\href
  {https://doi.org/10.1103/PhysRevA.87.051607} {\bibfield  {journal} {\bibinfo
  {journal} {Phys. Rev. A}\ }\textbf {\bibinfo {volume} {87}},\ \bibinfo
  {pages} {051607} (\bibinfo {year} {2013})}\BibitemShut {NoStop}%
\bibitem [{\citenamefont {Cataldo}\ and\ \citenamefont
  {Jezek}(2011)}]{Jezek-Bose-Hubbard-ring-lattice}%
  \BibitemOpen
  \bibfield  {author} {\bibinfo {author} {\bibfnamefont {H.~M.}\ \bibnamefont
  {Cataldo}}\ and\ \bibinfo {author} {\bibfnamefont {D.~M.}\ \bibnamefont
  {Jezek}},\ }\bibfield  {title} {\bibinfo {title} {Bose-hubbard model in a
  ring-shaped optical lattice with high filling factors},\ }\href
  {https://doi.org/10.1103/PhysRevA.84.013602} {\bibfield  {journal} {\bibinfo
  {journal} {Phys. Rev. A}\ }\textbf {\bibinfo {volume} {84}},\ \bibinfo
  {pages} {013602} (\bibinfo {year} {2011})}\BibitemShut {NoStop}%
\bibitem [{\citenamefont {Jezek}\ and\ \citenamefont
  {Cataldo}(2011)}]{Jezek-winding-number}%
  \BibitemOpen
  \bibfield  {author} {\bibinfo {author} {\bibfnamefont {D.~M.}\ \bibnamefont
  {Jezek}}\ and\ \bibinfo {author} {\bibfnamefont {H.~M.}\ \bibnamefont
  {Cataldo}},\ }\bibfield  {title} {\bibinfo {title} {Winding-number dependence
  of Bose-Einstein condensates in a ring-shaped lattice},\ }\href
  {https://doi.org/10.1103/PhysRevA.83.013629} {\bibfield  {journal} {\bibinfo
  {journal} {Phys. Rev. A}\ }\textbf {\bibinfo {volume} {83}},\ \bibinfo
  {pages} {013629} (\bibinfo {year} {2011})}\BibitemShut {NoStop}%
\bibitem [{\citenamefont {Arwas}\ and\ \citenamefont
  {Cohen}(2016)}]{Doron-Cohen-1}%
  \BibitemOpen
  \bibfield  {author} {\bibinfo {author} {\bibfnamefont {G.}~\bibnamefont
  {Arwas}}\ and\ \bibinfo {author} {\bibfnamefont {D.}~\bibnamefont {Cohen}},\
  }\bibfield  {title} {\bibinfo {title} {Chaos and two-level dynamics of the
  atomtronic quantum interference device},\ }\href@noop {} {\bibfield
  {journal} {\bibinfo  {journal} {New Journal of Physics}\ }\textbf {\bibinfo
  {volume} {18}},\ \bibinfo {pages} {015007} (\bibinfo {year}
  {2016})}\BibitemShut {NoStop}%
\bibitem [{\citenamefont {Arwas}\ \emph {et~al.}(2017)\citenamefont {Arwas},
  \citenamefont {Cohen}, \citenamefont {Hekking},\ and\ \citenamefont
  {Minguzzi}}]{Minguzzi-resonant-persistent}%
  \BibitemOpen
  \bibfield  {author} {\bibinfo {author} {\bibfnamefont {G.}~\bibnamefont
  {Arwas}}, \bibinfo {author} {\bibfnamefont {D.}~\bibnamefont {Cohen}},
  \bibinfo {author} {\bibfnamefont {F.}~\bibnamefont {Hekking}},\ and\ \bibinfo
  {author} {\bibfnamefont {A.}~\bibnamefont {Minguzzi}},\ }\bibfield  {title}
  {\bibinfo {title} {Resonant persistent currents for ultracold bosons on a
  lattice ring},\ }\href {https://doi.org/10.1103/PhysRevA.96.063616}
  {\bibfield  {journal} {\bibinfo  {journal} {Phys. Rev. A}\ }\textbf {\bibinfo
  {volume} {96}},\ \bibinfo {pages} {063616} (\bibinfo {year}
  {2017})}\BibitemShut {NoStop}%
\bibitem [{\citenamefont {Polo}\ \emph {et~al.}(2020)\citenamefont {Polo},
  \citenamefont {Naldesi}, \citenamefont {Minguzzi},\ and\ \citenamefont
  {Amico}}]{Minguzzi-PRA-two-bosons}%
  \BibitemOpen
  \bibfield  {author} {\bibinfo {author} {\bibfnamefont {J.}~\bibnamefont
  {Polo}}, \bibinfo {author} {\bibfnamefont {P.}~\bibnamefont {Naldesi}},
  \bibinfo {author} {\bibfnamefont {A.}~\bibnamefont {Minguzzi}},\ and\
  \bibinfo {author} {\bibfnamefont {L.}~\bibnamefont {Amico}},\ }\bibfield
  {title} {\bibinfo {title} {Exact results for persistent currents of two
  bosons in a ring lattice},\ }\href
  {https://doi.org/10.1103/PhysRevA.101.043418} {\bibfield  {journal} {\bibinfo
   {journal} {Phys. Rev. A}\ }\textbf {\bibinfo {volume} {101}},\ \bibinfo
  {pages} {043418} (\bibinfo {year} {2020})}\BibitemShut {NoStop}%
\bibitem [{\citenamefont {Richaud}\ and\ \citenamefont {Penna}(2019)}]{Penna}%
  \BibitemOpen
  \bibfield  {author} {\bibinfo {author} {\bibfnamefont {A.}~\bibnamefont
  {Richaud}}\ and\ \bibinfo {author} {\bibfnamefont {V.}~\bibnamefont
  {Penna}},\ }\bibfield  {title} {\bibinfo {title} {Pathway toward the
  formation of supermixed states in ultracold boson mixtures loaded in ring
  lattices},\ }\href {https://doi.org/10.1103/PhysRevA.100.013609} {\bibfield
  {journal} {\bibinfo  {journal} {Phys. Rev. A}\ }\textbf {\bibinfo {volume}
  {100}},\ \bibinfo {pages} {013609} (\bibinfo {year} {2019})}\BibitemShut
  {NoStop}%
\bibitem [{\citenamefont {Mu\~noz Mateo}\ \emph {et~al.}(2019)\citenamefont
  {Mu\~noz Mateo}, \citenamefont {Delgado}, \citenamefont {Guilleumas},
  \citenamefont {Mayol},\ and\ \citenamefont
  {Brand}}]{Guilleumas-nonlinear_ring}%
  \BibitemOpen
  \bibfield  {author} {\bibinfo {author} {\bibfnamefont {A.}~\bibnamefont
  {Mu\~noz Mateo}}, \bibinfo {author} {\bibfnamefont {V.}~\bibnamefont
  {Delgado}}, \bibinfo {author} {\bibfnamefont {M.}~\bibnamefont {Guilleumas}},
  \bibinfo {author} {\bibfnamefont {R.}~\bibnamefont {Mayol}},\ and\ \bibinfo
  {author} {\bibfnamefont {J.}~\bibnamefont {Brand}},\ }\bibfield  {title}
  {\bibinfo {title} {Nonlinear waves of Bose-Einstein condensates in rotating
  ring-lattice potentials},\ }\href
  {https://doi.org/10.1103/PhysRevA.99.023630} {\bibfield  {journal} {\bibinfo
  {journal} {Phys. Rev. A}\ }\textbf {\bibinfo {volume} {99}},\ \bibinfo
  {pages} {023630} (\bibinfo {year} {2019})}\BibitemShut {NoStop}%
\bibitem [{\citenamefont {Nigro}\ \emph {et~al.}(2018)\citenamefont {Nigro},
  \citenamefont {Capuzzi},\ and\ \citenamefont {Jezek}}]{Nigro_2018}%
  \BibitemOpen
  \bibfield  {author} {\bibinfo {author} {\bibfnamefont {M.}~\bibnamefont
  {Nigro}}, \bibinfo {author} {\bibfnamefont {P.}~\bibnamefont {Capuzzi}},\
  and\ \bibinfo {author} {\bibfnamefont {D.~M.}\ \bibnamefont {Jezek}},\
  }\bibfield  {title} {\bibinfo {title} {Blocked populations in ring-shaped
  optical lattices},\ }\href {https://doi.org/10.1103/PhysRevA.98.063622}
  {\bibfield  {journal} {\bibinfo  {journal} {Phys. Rev. A}\ }\textbf {\bibinfo
  {volume} {98}},\ \bibinfo {pages} {063622} (\bibinfo {year}
  {2018})}\BibitemShut {NoStop}%
\bibitem [{\citenamefont {Opatrn\'y}\ \emph {et~al.}(2015)\citenamefont
  {Opatrn\'y}, \citenamefont {Kol\'a\v{r}},\ and\ \citenamefont
  {Das}}]{Opatrny-Kolar-Das-LMG}%
  \BibitemOpen
  \bibfield  {author} {\bibinfo {author} {\bibfnamefont {T.}~\bibnamefont
  {Opatrn\'y}}, \bibinfo {author} {\bibfnamefont {M.}~\bibnamefont
  {Kol\'a\v{r}}},\ and\ \bibinfo {author} {\bibfnamefont {K.~K.}\ \bibnamefont
  {Das}},\ }\bibfield  {title} {\bibinfo {title} {Spin squeezing by tensor
  twisting and lipkin-meshkov-glick dynamics in a toroidal Bose-Einstein
  condensate with spatially modulated nonlinearity},\ }\href
  {https://doi.org/10.1103/PhysRevA.91.053612} {\bibfield  {journal} {\bibinfo
  {journal} {Phys. Rev. A}\ }\textbf {\bibinfo {volume} {91}},\ \bibinfo
  {pages} {053612} (\bibinfo {year} {2015})}\BibitemShut {NoStop}%
\bibitem [{\citenamefont {Kol\'a\ifmmode~\check{r}\else \v{r}\fi{}}\ \emph
  {et~al.}(2015)\citenamefont {Kol\'a\ifmmode~\check{r}\else \v{r}\fi{}},
  \citenamefont {Opatrn\'y},\ and\ \citenamefont
  {Das}}]{Opatrny-Kolar-Das-rotation}%
  \BibitemOpen
  \bibfield  {author} {\bibinfo {author} {\bibfnamefont {M.}~\bibnamefont
  {Kol\'a\ifmmode~\check{r}\else \v{r}\fi{}}}, \bibinfo {author} {\bibfnamefont
  {T.}~\bibnamefont {Opatrn\'y}},\ and\ \bibinfo {author} {\bibfnamefont
  {K.~K.}\ \bibnamefont {Das}},\ }\bibfield  {title} {\bibinfo {title}
  {Criticality and spin squeezing in the rotational dynamics of a Bose-Einstein
  condensate on a ring lattice},\ }\href
  {https://doi.org/10.1103/PhysRevA.92.043630} {\bibfield  {journal} {\bibinfo
  {journal} {Phys. Rev. A}\ }\textbf {\bibinfo {volume} {92}},\ \bibinfo
  {pages} {043630} (\bibinfo {year} {2015})}\BibitemShut {NoStop}%
\bibitem [{\citenamefont {Egorov}\ \emph {et~al.}(2013)\citenamefont {Egorov},
  \citenamefont {Opanchuk}, \citenamefont {Drummond}, \citenamefont {Hall},
  \citenamefont {Hannaford},\ and\ \citenamefont {Sidorov}}]{Egorov_Rb}%
  \BibitemOpen
  \bibfield  {author} {\bibinfo {author} {\bibfnamefont {M.}~\bibnamefont
  {Egorov}}, \bibinfo {author} {\bibfnamefont {B.}~\bibnamefont {Opanchuk}},
  \bibinfo {author} {\bibfnamefont {P.}~\bibnamefont {Drummond}}, \bibinfo
  {author} {\bibfnamefont {B.~V.}\ \bibnamefont {Hall}}, \bibinfo {author}
  {\bibfnamefont {P.}~\bibnamefont {Hannaford}},\ and\ \bibinfo {author}
  {\bibfnamefont {A.~I.}\ \bibnamefont {Sidorov}},\ }\bibfield  {title}
  {\bibinfo {title} {Measurement of $s$-wave scattering lengths in a
  two-component Bose-Einstein condensate},\ }\href
  {https://doi.org/10.1103/PhysRevA.87.053614} {\bibfield  {journal} {\bibinfo
  {journal} {Phys. Rev. A}\ }\textbf {\bibinfo {volume} {87}},\ \bibinfo
  {pages} {053614} (\bibinfo {year} {2013})}\BibitemShut {NoStop}%
\bibitem [{\citenamefont {Bennett}\ \emph {et~al.}(1996)\citenamefont
  {Bennett}, \citenamefont {Bernstein}, \citenamefont {Popescu},\ and\
  \citenamefont {Schumacher}}]{Bennett_entanglement}%
  \BibitemOpen
  \bibfield  {author} {\bibinfo {author} {\bibfnamefont {C.~H.}\ \bibnamefont
  {Bennett}}, \bibinfo {author} {\bibfnamefont {H.~J.}\ \bibnamefont
  {Bernstein}}, \bibinfo {author} {\bibfnamefont {S.}~\bibnamefont {Popescu}},\
  and\ \bibinfo {author} {\bibfnamefont {B.}~\bibnamefont {Schumacher}},\
  }\bibfield  {title} {\bibinfo {title} {Concentrating partial entanglement by
  local operations},\ }\href {https://doi.org/10.1103/PhysRevA.53.2046}
  {\bibfield  {journal} {\bibinfo  {journal} {Phys. Rev. A}\ }\textbf {\bibinfo
  {volume} {53}},\ \bibinfo {pages} {2046} (\bibinfo {year}
  {1996})}\BibitemShut {NoStop}%
\bibitem [{\citenamefont {Leonhardt}(2010)}]{leonhardt_2010}%
  \BibitemOpen
  \bibfield  {author} {\bibinfo {author} {\bibfnamefont {U.}~\bibnamefont
  {Leonhardt}},\ }\href@noop {} {\emph {\bibinfo {title} {Essential Quantum
  Optics: From Quantum Measurements to Black Holes}}}\ (\bibinfo  {publisher}
  {Cambridge University Press},\ \bibinfo {year} {2010})\ pp.\ \bibinfo {pages}
  {92--134}\BibitemShut {NoStop}%
\bibitem [{\citenamefont {Yurke}\ \emph {et~al.}(1986)\citenamefont {Yurke},
  \citenamefont {McCall},\ and\ \citenamefont {Klauder}}]{Klauder}%
  \BibitemOpen
  \bibfield  {author} {\bibinfo {author} {\bibfnamefont {B.}~\bibnamefont
  {Yurke}}, \bibinfo {author} {\bibfnamefont {S.~L.}\ \bibnamefont {McCall}},\
  and\ \bibinfo {author} {\bibfnamefont {J.~R.}\ \bibnamefont {Klauder}},\
  }\bibfield  {title} {\bibinfo {title} {SU(2) and SU(1,1) interferometers},\
  }\href {https://doi.org/10.1103/PhysRevA.33.4033} {\bibfield  {journal}
  {\bibinfo  {journal} {Phys. Rev. A}\ }\textbf {\bibinfo {volume} {33}},\
  \bibinfo {pages} {4033} (\bibinfo {year} {1986})}\BibitemShut {NoStop}%
\bibitem [{\citenamefont {Cejnar}\ \emph {et~al.}(2021)\citenamefont {Cejnar},
  \citenamefont {Str\'{a}nsk\'{y}}, \citenamefont {Macek},\ and\ \citenamefont
  {Kloc}}]{Cejnar_2021}%
  \BibitemOpen
  \bibfield  {author} {\bibinfo {author} {\bibfnamefont {P.}~\bibnamefont
  {Cejnar}}, \bibinfo {author} {\bibfnamefont {P.}~\bibnamefont
  {Str\'{a}nsk\'{y}}}, \bibinfo {author} {\bibfnamefont {M.}~\bibnamefont
  {Macek}},\ and\ \bibinfo {author} {\bibfnamefont {M.}~\bibnamefont {Kloc}},\
  }\bibfield  {title} {\bibinfo {title} {Excited-state quantum phase
  transitions},\ }\href@noop {} {\bibfield  {journal} {\bibinfo  {journal}
  {Journal of Physics A: Mathematical and Theoretical}\ }\textbf {\bibinfo
  {volume} {54}},\ \bibinfo {pages} {133001} (\bibinfo {year}
  {2021})}\BibitemShut {NoStop}%
\bibitem [{\citenamefont {Bennett}\ \emph {et~al.}(1993)\citenamefont
  {Bennett}, \citenamefont {Brassard}, \citenamefont {Cr\'epeau}, \citenamefont
  {Jozsa}, \citenamefont {Peres},\ and\ \citenamefont
  {Wootters}}]{Wootters_teleportation}%
  \BibitemOpen
  \bibfield  {author} {\bibinfo {author} {\bibfnamefont {C.~H.}\ \bibnamefont
  {Bennett}}, \bibinfo {author} {\bibfnamefont {G.}~\bibnamefont {Brassard}},
  \bibinfo {author} {\bibfnamefont {C.}~\bibnamefont {Cr\'epeau}}, \bibinfo
  {author} {\bibfnamefont {R.}~\bibnamefont {Jozsa}}, \bibinfo {author}
  {\bibfnamefont {A.}~\bibnamefont {Peres}},\ and\ \bibinfo {author}
  {\bibfnamefont {W.~K.}\ \bibnamefont {Wootters}},\ }\bibfield  {title}
  {\bibinfo {title} {Teleporting an unknown quantum state via dual classical
  and Einstein-podolsky-rosen channels},\ }\href
  {https://doi.org/10.1103/PhysRevLett.70.1895} {\bibfield  {journal} {\bibinfo
   {journal} {Phys. Rev. Lett.}\ }\textbf {\bibinfo {volume} {70}},\ \bibinfo
  {pages} {1895} (\bibinfo {year} {1993})}\BibitemShut {NoStop}%
\bibitem [{\citenamefont {Braunstein}\ and\ \citenamefont
  {Kimble}(1998)}]{Braunstein_Kimble}%
  \BibitemOpen
  \bibfield  {author} {\bibinfo {author} {\bibfnamefont {S.~L.}\ \bibnamefont
  {Braunstein}}\ and\ \bibinfo {author} {\bibfnamefont {H.~J.}\ \bibnamefont
  {Kimble}},\ }\bibfield  {title} {\bibinfo {title} {Teleportation of
  continuous quantum variables},\ }\href
  {https://doi.org/10.1103/PhysRevLett.80.869} {\bibfield  {journal} {\bibinfo
  {journal} {Phys. Rev. Lett.}\ }\textbf {\bibinfo {volume} {80}},\ \bibinfo
  {pages} {869} (\bibinfo {year} {1998})}\BibitemShut {NoStop}%
\end{thebibliography}

%

\vfill

\end{document}